\documentclass[dvips]{article}
\usepackage{supertabular,lscape,epsfig}
\usepackage{amssymb}
\usepackage{amsmath}

\DeclareSymbolFont{ppa}{OT1}{ppl}{m}{it}
\DeclareMathSymbol{\vv}{\mathalpha}{ppa}{'166}

\begin{document}

\newcommand{\dd}{\,{\rm d}}
\newcommand{\ie}{{\it i.e.},\,}
\newcommand{\etal}{{\it et al.\ }}
\newcommand{\eg}{{\it e.g.},\,}
\newcommand{\cf}{{\it cf.\ }}
\newcommand{\vs}{{\it vs.\ }}
\newcommand{\zdot}{\makebox[0pt][l]{.}}
\newcommand{\up}[1]{\ifmmode^{\rm #1}\else$^{\rm #1}$\fi}
\newcommand{\dn}[1]{\ifmmode_{\rm #1}\else$_{\rm #1}$\fi}
\newcommand{\upd}{\up{d}}
\newcommand{\uph}{\up{h}}
\newcommand{\upm}{\up{m}}
\newcommand{\ups}{\up{s}}
\newcommand{\arcd}{\ifmmode^{\circ}\else$^{\circ}$\fi}
\newcommand{\arcm}{\ifmmode{'}\else$'$\fi}
\newcommand{\arcs}{\ifmmode{''}\else$''$\fi}
\newcommand{\MS}{{\rm M}\ifmmode_{\odot}\else$_{\odot}$\fi}
\newcommand{\RS}{{\rm R}\ifmmode_{\odot}\else$_{\odot}$\fi}
\newcommand{\LS}{{\rm L}\ifmmode_{\odot}\else$_{\odot}$\fi}

\newcommand{\Abstract}[2]{{\footnotesize\begin{center}ABSTRACT\end{center}
\vspace{1mm}\par#1\par
\noindent
{~}{\it #2}}}

\newcommand{\TabCap}[2]{\begin{center}\parbox[t]{#1}{\begin{center}
  \small {\spaceskip 2pt plus 1pt minus 1pt T a b l e}
  \refstepcounter{table}\thetable \\[2mm]
  \footnotesize #2 \end{center}}\end{center}}

\newcommand{\TableSep}[2]{\begin{table}[p]\vspace{#1}
\TabCap{#2}\end{table}}

\newcommand{\FigCap}[1]{\footnotesize\par\noindent Fig.\  %
  \refstepcounter{figure}\thefigure. #1\par}

\newcommand{\TableFont}{\footnotesize}
\newcommand{\TableFontIt}{\ttit}
\newcommand{\SetTableFont}[1]{\renewcommand{\TableFont}{#1}}

\newcommand{\MakeTable}[4]{\begin{table}[htb]\TabCap{#2}{#3}
  \begin{center} \TableFont \begin{tabular}{#1} #4 
  \end{tabular}\end{center}\end{table}}

\newcommand{\MakeTableSep}[4]{\begin{table}[p]\TabCap{#2}{#3}
  \begin{center} \TableFont \begin{tabular}{#1} #4 
  \end{tabular}\end{center}\end{table}}

\newenvironment{references}%
{
\footnotesize \frenchspacing
\renewcommand{\thesection}{}
\renewcommand{\in}{{\rm in }}
\renewcommand{\AA}{Astron.\ Astrophys.}
\newcommand{\AAS}{Astron.~Astrophys.~Suppl.~Ser.}
\newcommand{\ApJ}{Astrophys.\ J.}
\newcommand{\ApJS}{Astrophys.\ J.~Suppl.~Ser.}
\newcommand{\ApJL}{Astrophys.\ J.~Letters}
\newcommand{\AJ}{Astron.\ J.}
\newcommand{\IBVS}{IBVS}
\newcommand{\PASP}{P.A.S.P.}
\newcommand{\Acta}{Acta Astron.}
\newcommand{\MNRAS}{MNRAS}
\renewcommand{\and}{{\rm and }}
\section{{\rm REFERENCES}}
\sloppy \hyphenpenalty10000
\begin{list}{}{\leftmargin1cm\listparindent-1cm
\itemindent\listparindent\parsep0pt\itemsep0pt}}%
{\end{list}\vspace{2mm}}

\def\TYLDA{~}
\newlength{\DW}
\settowidth{\DW}{0}
\newcommand{\dw}{\hspace{\DW}}

\newcommand{\refitem}[5]{\item[]{#1} #2%
\def\REFARG{#3}\ifx\REFARG\TYLDA\else, {\it#3}\fi
\def\REFARG{#4}\ifx\REFARG\TYLDA\else, {\bf#4}\fi
\def\REFARG{#5}\ifx\REFARG\TYLDA\else, {#5}\fi.}

\newcommand{\Section}[1]{\section{\hskip-6mm.\hskip3mm#1}}
\newcommand{\Subsection}[1]{\subsection{#1}}
\newcommand{\Acknow}[1]{\par\vspace{5mm}{\bf Acknowledgements.} #1}
\pagestyle{myheadings}

\newfont{\bb}{ptmbi8t at 12pt}
\newcommand{\xrule}{\rule{0pt}{2.5ex}}
\newcommand{\xxrule}{\rule[-1.8ex]{0pt}{4.5ex}}
\def\thefootnote{\fnsymbol{footnote}}
\begin{center}
{\Large\bf Gravitational Microlensing:\\
 Black Holes, Planets;\\
\vskip2mm
OGLE, VLTI, HST and Space Probes}
\vskip.6cm
{\bf
B.~~P~a~c~z~y~{\'n}~s~k~i}
\vskip2mm
{Princeton University Observatory, Princeton, NJ 08544-1001, USA\\
e-mail: bp@astro.princeton.edu}
\end{center}

\Abstract{
OGLE and other projects are likely to discover first stellar
mass black holes and the first planets through gravitational lensing
in the next year or two.  It is important to have follow-up projects
ready, using diverse observing methods.  The best for black hole 
detection would be a measurement of image splitting with VLTI, or
any other optical interferometer.  Alternative approach is to measure
non-linear motion of the light centroid with the HST, or even with
a ground based telescope.  Every year OGLE detects several very long
duration microlensing events brighter than $ I = 16 $ mag and $ K = 14 $ mag.
The two images may be separated by up to 10 mas.

Ground based detection of strong caustic
crossing planetary events will provide mass ratios and proper motions
for the detected systems.  For most events photometric parallax
needed for mass determination will require a space instrument at 
least as far as the L2 point, to provide long enough baseline.
}{~}

\noindent
{\bf Key words:}{\it Astrometry- Black Holes - Dark matter - 
Gravitational lensing -Planets}

\Section{Introduction}

The main purpose of this paper is to bring to reader's attention the
current status of microlensing searches.  It is likely that
within a year or two these searches will lead to firm discoveries
of stellar mass black holes and planets.
The paper is centered on current developments at the
Optical Gravitational Lensing Experiment (OGLE, Udalski et al. 1997, 2002).
It is motivated by the fact that OGLE team is small and hence not in a
position to do the diverse follow-up observations, which are needed to
make conclusive case for black hole lenses, or to measure planetary 
masses.  This is one of the reasons the
OGLE data are made public domain with no strings attached, so other teams
or individuals may use them as the basis for their own follow-up projects.

The search for gravitational microlensing, and follow-up observations,
are a well developed `industry', with dozens of events reported every year
in real time by the MOA collaboration (Bond et al. 2001):

\centerline{http://www.roe.ac.uk/\%7Eiab/alert/alert.html}

\noindent
and hundreds reported by the OGLE collaboration (Udalski et al. 1997, 2002):

\centerline{http://www.astrouw.edu.pl/\~~ogle/ogle3/ews/ews.html}

\noindent
A major increase in the discovery rate is to be expected when the
upgrade of MOA capability is completed, with the first light from
the new 1.8 meter telescope expected in 2004 or 2005:

\centerline{http://www3.vuw.ac.nz/scps/moa/}

\noindent
While there is a very large diversity of research topics related to 
microlensing, as well as very many other results of the surveys (cf.
Paczy\'nski 1996a, Gould 2001, and references therein), this
paper is focused on just two extreme cases: the search for black holes
and the search for planets, i.e. very large and very small mass
lenses.  A definite detection of stellar mass black holes and planets
is an interesting goal, almost certainly achievable within 
a year or two.  Only microlensing is capable of detecting Earth mass
planets and isolated black holes  with the technology which already exists.
This paper outlines OGLE contribution to the task.  However, to be
successful it requires diverse follow-up observations to be carried out
by astronomers not necessarily associated with OGLE.

\Section{Black Holes}

The discovery of long duration microlensing events prompted suggestions
that some of the lenses may be due to stellar mass black holes (Bennett
et al. 2002a,b; Mao et al. 2002; Smith 2003).  The argument is simple: the
time scale of a microlensing event, i.e. the time $ t_E $ it takes a lens 
to move with respect to the source by the Einstein radius $ \varphi _E $
is given as (cf. Paczy\'nski 1996a, and references therein)
$$
 t_E = { \varphi _E \over \dot \varphi } = 1.01 ~ yr ~
 \left( { M \over M_{\odot} } \right) ^{1/2} 
 \left( { 8 ~ kpc \over D } \right) ^{1/2}
 \left( { 1 ~ mas ~ yr^{-1} \over \dot \varphi } \right) ,
\eqno(1)
$$
where $ \dot \varphi $ is a relative proper motion of a lens with respect 
to the source, $ M $ is the lens mass, and $ D $ is the effective distance
defined as
$$
D = { D_s D_d \over D_s - D_d } ,
\eqno(2)
$$
where $ D_s $ is the distance to the source, and $ D_d $ is the distance
to the lensing object (deflector).  The angular Einstein radius is given as
$$
\varphi _E = 1.01 ~ {\rm mas} ~ 
\left( { M \over M_{\odot} } \right) ^{1/2} 
\left( { 8 ~ kpc \over D } \right) ^{1/2} .
\eqno(3)
$$

In most cases the distance to the source is either known or may be
estimated, but the distance to the lens, its mass, and its proper motion
with respect to the source, are not known.  The time scale of a
microlensing event, $ t_E $, is directly measured, and the longer it
is the larger the lens mass, other things being equal.  Unfortunately,
other things ($ D_d $ and $ \dot \varphi $) are not equal and may very
a lot from one event to another.  Therefore, a long duration is only
an indication of a large mass, but only in a statistical sense.

For long events the effect of Earth's motion around the sun makes the
relative motion in the observer -- lens -- source system not linear
(Refsdal 1966, Gould 1992), and leads to the so called photometric parallax
effect, which is always measured for long events.  The most
spectacular case known so far is that of OGLE-1999-BUL-19 (Smith et al.
2002), in which several maxima were observed as a consequence of the
Earth's orbital motion.  The amplitude of the parallax effect provides
additional information, and partly removes the degeneracy in the eq. (1).
The smaller the relative proper motion $ \dot \varphi $, the stronger the
effect.  In the case of OGLE-1999-BUL-19 it was clear that the relative
proper motion $ \dot \varphi $
was very small, and that small velocity was responsible for the long
time scale, $ t_E = 1.02 $ years, while the lens mass was sub-solar.

More interesting was OGLE-1999-BUL-32 = MACHO-99-BLG-22
(Mao et al. 2002, Bennett et al. 2002b), the longest event so far, with
$ t_E = 1.75 $ years, and clearly measured, but small parallax effect,
indicative of a large relative proper motion, and hence a large lens mass.
This is currently the best candidate for gravitational microlensing 
due to stellar mass black hole, but the mass estimates requires 
statistical analysis, which is not very reliable.  In fact the
probability estimate that the lens is a black hole changed from 20\% to 76\%
between v1 and v4 edition of astro-ph/0203257 (Agol et al. 2002).  It is
not possible to make a credible probability estimate as the stellar mass
function, all the way from brown dwarfs to intermediate mass black holes,
is not known.

The second strong black hole candidate is OGLE-SC5\_2859 (Wo\'zniak et al.
2001).  With $ t_E = 1.5 $ years this is the second longest event known.
The parallax effect is clearly detected, but its amplitude is small.
However, there may be other complications (Smith 2003),

Unfortunately, in all cases so far the third quantity which is in principle
observable: $ \varphi _E $, remains unknown.  In fact there is not a single
microlensing event for which Einstein radius was directly measured.  The 
reason is simple: the expected value of $ \varphi _E $ is of the order of
a milli arc second (Eq. 2), well 
below the resolution of existing optical instruments.  There is only one
case, EROS BLG-2000-5, a binary lens, for which it was possible 
to infer $ \varphi _ E = 1.4 $ mas as the proper motion was measured with
caustic crossing, and photometric parallax effect was also detected
(An et al. 2002).  The total mass of the lens, and the mass ratio, were
found to be $ 0.61 ~ M_{\odot} $, and $ M_2/M_1 = 0.748 $, respectively.

\Subsection{Resolving Microlensing - VLTI}

A direct method to measure $ \varphi _E $ was proposed
by Delplancke et al. (2001): to resolve the double image generated by 
microlensing with an optical interferometer, specifically VLTI.
The first papers with the first results from VLTI are already
published (Segransan et al. 2003), or posted (Kervella et al. 2003),
and perhaps VLTI will soon be used to measure microlensed stars.

For an event well covered photometrically the dimensionless angular
separation between the source and the lens is known as a function of time:
$$
u = { \varphi \over \varphi _E } =
\left[ u_{min}^2 + \left( { t - t_{max} \over t_E } \right) ^2 \right] ^{1/2} ,
\eqno(4)
$$
where $ \varphi $ is the angular separation between the lens and the source.
The combined brightness of the two images is larger than the source
by the so called magnification factor A:
$$
A = { u^2 +2 \over u \sqrt{ u^2 + 4 } } .
\eqno(5)
$$
The magnification is the largest when the separation is the smallest,
i.e. when $ u = u_{min} $, which happens at the time $ t = t_{max} $.
Photometric model of a well observed microlensing event provides accurate 
values of $ u_{min} $, $ t_{max} $, and $ t_E $.

The two images have opposite parity: one positive, with magnification $ A_+ $,
always located outside of Einstein circle, another negative, with magnification
$ A_- $, always located within Einstein circle.  We have
$$
A_+ + A_- = A , \hskip 1.0cm A_+ - A_- = 1 .
\eqno(6)
$$
The two images are separated by an angle $ \Delta \varphi _{+,-} $ given as
$$
{ \Delta \varphi _{+,-} \over \varphi _E } = \sqrt{ u^2 + 4 } .
\eqno(7)
$$

A single measurement of the angular separation $ \Delta \varphi _{+,-} $
between the two micro-images provides a direct determination of angular
Einstein radius, as $ u(t) $ is provided by the photometric model of
the event.  While such a measurement is very robust and elegant, the VLTI
(or any other optical interferometer) has to be able to observe stars
as faint as those which are microlensed.  Several microlensed stars
are bright, with $ I \sim 16 $ mag, and $ K \sim 14 $ mag.  At peak
magnification some reach $ I \approx 14 $ mag, i.e. $ K \approx 12 $ mag.
Bright, long duration OGLE events are described in the Section: 
{\it Long Current Events}.

\Subsection{Astrometric Shift - HST}

An alternative astrometric approach is not to resolve the double
image, but to measure the motion of the light centroid (Hog et al. 1995, 
Miyamoto and Yoshi 1995, Walker 1995, Boden et al. 1998, Paczy\'nski 1998).
In all these papers it was suggested that a space probe, GAIA or SIM:

\centerline{http://astro.estec.esa.nl/GAIA/}

\centerline{http://sim.jpl.nasa.gov/whatis/}

\noindent
will be able to do micro arc second precision astrometry, and therefore
will be able to determine angular Einstein radii for virtually any microlensing
event.

The light centroid shifts with respect to the source position by
$$
\delta \varphi = { u \over u^2 + 2 } ~ \varphi _E ,
\eqno(8)
$$
The displacement reaches maximum value $ \delta \varphi _{max} $
$$
\delta \varphi _{max}  = 2^{-3/2} ~ \varphi _E \approx 0.354 ~ \varphi _E 
\hskip 0.5cm {\rm for} \hskip 0.5cm u_{max} = 2^{1/2} ~ \approx 1.414 .
\eqno(9)
$$
This angular separation corresponds to a magnification
$ A = 1.155 $, i.e. $ 0.156 $ mag above the baseline.
The full range of displacements, from the moment when $ A = 1.155 $
on the rising branch, to the moment when $ A = 1.155 $ on the descending
branch, is $ t_E \times 2 \times 2^{1/2} \approx 2.8 t_E $.  This is
inconvenient, as we are targeting long duration events, with $ t_E > 1 $ year.

\begin{figure}[t]  
\vspace{9.0cm}
\includegraphics{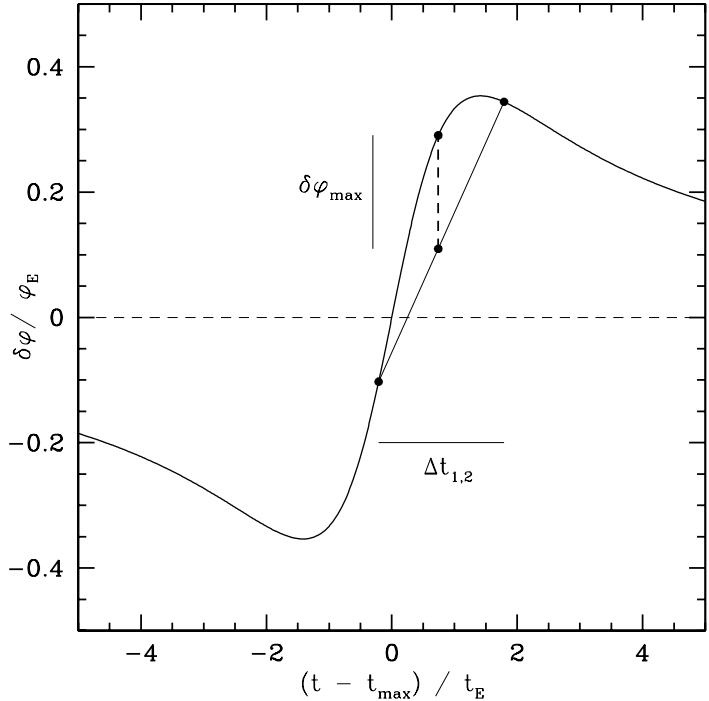}
\caption{
The motion of the light centroid is shown as a function of time for a simple
case, when the impact parameter for microlensing is exactly zero.
For a given value of observing time interval, $ \Delta t_{1,2} $, there
is an optimum choice of the beginning and the end of observing, $ t_1 $ and
$ t_2 $, respectively, which maximizes the nonlinearity of proper motion,
$ \delta \varphi _{max} $.  An example shown corresponds to
$ \Delta t_{1,2} = 2 t_E $.
}
\end{figure}

While SIM and GAIA will be able to do a superb job in measuring the shifts
in the light centroids of almost all selected microlensing events, they
are scheduled for launch in 2009 and in $ 2010 - 2012 $, respectively.  This
is a long time to wait.  In this paper I am most interested in a possibility
of identifying black hole lenses, and in measuring their masses in the next
year or two.  The best candidates for such studies are the longest 
microlensing events, which are likely to have the largest Einstein radii, 
probably in the range of several milli arc seconds.
Small angle astrometry can be done with the HST with a precision of a fraction
of a milliarcseconds (cf. Fritz et al. 2002, and references therein).  Perhaps
modern large telescope can also achieve such precision.  The advantage of
this approach is that the brightness of the lensed star is not a problem, 
but many images covering a substantial time interval have to be
obtained.  As we are focusing on long duration events, the time interval
may be several years.

The following is an illustration of what may be expected.  For simplicity
I assume that the impact parameter of the lensing event is
very small, i.e. $ u_{min} \ll 1 $, and hence the maximum magnification
will be very large, $ A_{max} \gg 1 $ (cf. eqs. 4, 5).  This
implies that the source moves along a line in the sky
which passes almost exactly behind the lens, the relative proper motion
is one-dimensional, and so is the shift of the light centroid.  This
makes the presentation of motion very simple.  The time dependence of
centroid displacement with respect to the source is shown with solid line 
in Fig. 1.  The maximum range is $ 0.707 \varphi _E $, but it takes
time interval of $ 2.83 t_E $, as given with the eqs. (9).  Unfortunately,
we do not know the asymptotic motion of the source, unperturbed by lensing.
To learn this we would have to monitor it for a very long time indeed, as 
the solid line in Fig. 1 approaches the unperturbed source position very
slowly, as $ 1 / t $.  What can be measured within a reasonable time
interval is the non-linear motion
of the light centroid.  In most cases of interest the astrometric parallax
effect due to Earth orbital motion is negligible (Figs. 1 and 2 in 
Paczy\'nski 1998), so it is sufficient to calculate the departure from
linear motion based on Fig. 1.

\begin{figure}[t]  
\vspace{9.0cm}
\includegraphics{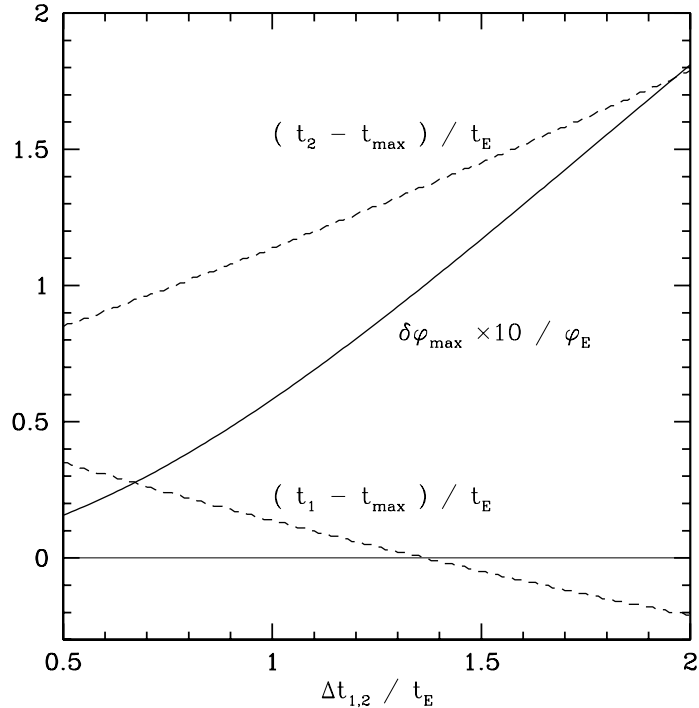}
\caption{
For a given time interval of observations, $ \Delta t_{1,2} = t_2 - t_1 $,
the nonlinearity of centroid proper motion $ \delta _{\varphi} $ depends
on the choice of $ t_1 $.  The figure shows the values of $ t_1 $
and $ t_2 $ which maximize $ \delta _{\varphi} $,
as a function of $ \Delta t_{1,2} $).
}
\end{figure}

Let us assume that astrometric measurements are done over time
interval $ \Delta t_{1,2} $, beginning at some time $ t_1 $, and 
continued till $ t_2 $.  For a given value of $ \Delta t_{1,2} $
we want to select the beginning of astrometric observations $ t_1 $,
to maximize the deviation of centroid position from a linear proper
motion.  For example, the two dots on the solid curve in Fig. 1,
connected with a straight solid line represent an example for
$ \Delta t_{1,2} = 2 t_E $.  The vertical dashed line represents the
maximum value of departure from linear motion, in this case it is 
$ 0.18 \varphi _E $.

More general results are shown in Fig. 2, where the beginning and the
end of observations, $ t_1 $ and $ t_2 $ are shown with dashed lines
as a function of $ \Delta t_{1,2} $, all in units of Einstein time
scale $ t_E $.  The maximum displacement with respect
to linear proper motion in the considered time interval is shown with
thick solid line, with the value of $ \delta \varphi _{max} $ multiplied
by a factor 10.  It is interesting that $ \delta \varphi _{max} $
is approximately a linear function of $ \Delta t_{1,2} $.

Fig. 2 presents a practical problem with astrometry of centroid shift
of long microlensing events: in order to detect a substantial 
non-linearity of centroid proper motion it is necessary to monitor
it for a long time.  For example, if $ t_E = 1.5 $ years, and 
$ \varphi _E = 5 $ mas, then monitoring the motion over 3 year time
interval provides us with the maximum displacement
$ \delta \varphi _{max} = 0.18 \times \varphi _E = 0.9 $ mas.
It is convenient that astrometry has to begin not much
earlier than the peak magnification.  By that time all 
photometric parameters of the microlensing event are fairly well known,

\Subsection{Long Current Events}

It is fortunate that to achieve the largest non-linear shift in the
light centroid there is no need to begin astrometry while the microlensing
event is not fully developed.  The OGLE EWS alert is usually issued when
the apparent magnitude of a candidate event is 0.2 -- 0.3 mag above the 
baseline.  At this early stage it is not always clear how long the event
is going to be, what will be its peak magnification, and even will it be 
a microlensing event at all.   As explained in previous section it is 
reasonable to begin accurate astrometry some time prior to the peak 
magnification.

In the past the longest microlensing events:
OGLE-1999-BUL-32, OGLE-SC5\_2859, and OGLE-1999-BUL-19, were all recognized
as very long well past their peak, analyzing archive data.  There were
two main reason for this time lag: improved photometric accuracy had to
be developed, using new image subtraction software (DIA - Difference Image 
Analysis: Alard and Lupton 1998, Alard 2000, Wo\'zniak 2002), and it had
to be recognize that events with a time scale well over a year existed.
The DIA is now used in real time data analysis, and it is known that 
very long events exist, so the task of early recognition of the very
long events is now easy.

The following is the current status of OGLE EWS results.
The unfolding OGLE-2003-BLG-192 has barely begun 
its rise, with the time scale estimate $ t_E \approx 235 $ days.
The lensed star is bright: $ {\rm I = 16 }$ mag, but the expected peak
magnification is low.  At this time it cannot be guaranteed that this is a
microlensing event.

There are several other long events just beginning their rise, 
with $ I \sim 16 $ mag, or brighter: OGLE-2003-BLG-047 with $ t_E = 156 $
days, OGLE-2003-BLG-188 with $ t = 121 $ days. OGLE-2002-BLG-360, with
$ t_E = 270 $ days, is close to its peak at $ I = 14.2 $.
Two bright and long
events are almost over: OGLE-2002-BLG-061 with $ t_E = 305 $ days, and
OGLE-2002-BLG-334 with $ t_E = 160 $ days.  Many more long events are
faint at their baseline, heavily blended and therefore not very useful.
However, several events every year are long and reasonably bright, with
$ I \le 16 $ mag.  This may be time to
propose follow-up astrometric observations for some of these events. 
Most lensed stars are in the Galactic Bulge and they are red,
with $ (I-K) \ge 2 $ mag, i.e. the bright events have $ K < 14 $ mag.

\Section{Planets}

Another high profile discovery which is likely to result from microlensing
is the first firm detection of a planetary signal, as proposed by Mao and 
Paczy\'nski (1991).  There is a rich literature on the subject, devoted
to various aspects of planetary lensing, and how to recognize it. 
As planetary events are even less probable than stellar, a lot of theoretical
effort was directed to increase the probability, at the cost of
seeking low amplitude signals.  This may be a good approach, but not
at the beginning of a search.  Right now we need credible
events with no ambiguity about their interpretation.
In other words, one needs an `overkill' evidence to be persuasive the
first time, and perhaps even the first several times.  Such clear
planetary signal will be present in caustic crossing events, like those
shown in Fig. 10 of Paczy\'nski (1996a).  Even Earth mass planet can
generate a complicated disturbance in a stellar microlensing light curve,
with an amplitude of 20\% or more.  Such events are far less frequent
than 5\% or 2\% amplitude planetary disturbances, but they offer unambiguous
evidence of a planet, and an unambiguous mass ratio.  Also, as any caustic
crossing event, they will provide the information about the relative proper
motion $ \dot \varphi $ (Gould 1994, Graff and Gould 2002).

While any strong planetary event is likely to provide the mass ratio and
a relative proper motion, it is necessary to measure also photometric
parallax effect to determine
the mass.  This may be possible in some cases, when the stellar
event time scale $ t_E $ is long, but in general this can be done only
with spacecraft observations (Gould 1995, Gould et al. 2003), as they
provide a baseline for the photometric parallax determination.
Following the analysis of a possible planetary signal from OGLE-2002-BLG-055
(Jaroszy\'nski and Paczy\'nski 2002) the EWS alert system on OGLE has been
upgraded to EEWS (Early Early Warning System), capable of recognizing
and verifying in real time  possible planetary disturbances in stellar 
microlensing events (A. Udalski, 2003, private communication).  No
structure is expected on a time scale shorter than $ \sim 1 $ hour, as
it takes so long for a source star to move by its own radius.  Therefore,
instant follow-up observations at $ \sim 30 $ minute intervals are planned
by the OGLE when a plausible planetary disturbance will be recognized.
The new system should be much more sensitive to planet detection that
the past attempts, which generated only upper limits (e.g. Gaudi et al. 2002).

The most natural follow-up observations are photometric measurements to
be done from other location, to provide full time coverage.  There
are several other active microlensing projects: MOA (Mond et al. 2001),
PLANET (Albrow et al. 2002), MPS (Rhie et al. 1999), GMAN (Becker et al.
1997, Bennett et al. 2002a), MicroFUN (A. Gould, 
2003, private communication)

\centerline{http://www.astronomy.ohio-state.edu/\~~microfun}

\noindent
This coverage should be adequate to obtain the mass ratio, a clear indicator
of a planetary presence.  Spacecraft observations are needed to determine
the masses, as well as the linear planet - star separations (in Astronomical
Units).

\Section{Conclusions}

It is very likely that stellar mass black holes and planets will be 
discovered in the next year or two.  OGLE is generating several
bright and long duration microlensing events every year (cf. Section 2.3),
and these are promising candidates for stellar mass black holes.  All
these events have not only their times scales $ t_E $ measured, but
also the photometric parallax effect.  To make
a definite case for black holes it is necessary to measure 
their angular Einstein radii, $ \varphi _E $.  This could be done
best with an optical interferometer, possibly with VLTI (Delplancke
et al. 2001).  An alternative approach is to carry out accurate astrometric
monitoring and to detect a non-linear motion of the light centroid with 
the HST, or perhaps
even with new large ground based telescopes.  A lens with 
$ \sim 10 ~ M_{\odot} $ is likely to have $ \varphi _E \sim 3 $ mas.
If intermediate mass black holes exist (Madau and Rees 2001,
Chisholm et al. 2002, Wu et al. 2002, and references therein) the corresponding
image splittings could be much larger, with $ \varphi _E \sim 10 $ mas,
and perhaps even more.  Such events would make image centroid motion
easily detectable not only with the HST, but also with ground based
telescopes.

Once several firm planetary detections are made we shall be in a much better
position to plan future upgrades of planetary microlensing searches.
It is likely that the detection threshold could be lowered to improve the
statistics without compromising credibility.  But a major improvement will 
require a major increase in the
photometric data rate, to monitor more stars more frequently and with
a higher photometric accuracy.  Note: vast majority of microlensing events
are heavily blended and are relatively faint.  It will take a lot of photons
to reliably search them for small planetary disturbances.

While OGLE and other ground based telescopes will determine the mass ratios
for star - planet systems, and the source - lens proper motion in
caustic crossing events, only some of them will have large enough
$ t_E $ to allow photometric parallax to be measured from the ground (cf. 
Jaroszy\'nski and Paczy\'nski 2002), and to determine the masses as well.
In most cases it will take space instrument to provide a large
baseline to measure photometric parallax effect, and hence the masses
for most of them (Gould 1995, Gould et al. 2003).

This paper is posted on astro-ph only.  Its purpose is to outline prospects
for various space and ground based projects which are needed to provide
definite evidence for stellar mass black holes and planets, using OGLE
as a search tool.

\Acknow{This research was supported by the NASA grant NAG5-12212 and the 
NSF grant AST-0204908.}


\end{document}